\documentclass{jpsj3}
\usepackage{graphicx}% Include figure files
\usepackage{dcolumn}%
\usepackage{bm,times,amsmath,amssymb}

\title{Size Dependence of Current--Voltage Properties\\in Coulomb Blockade Networks}

\author{
 \name{Takayuki \surname{Narumi} \thanks{E-mail address: narumi@athena.ap.kyushu-u.ac.jp}, 
 \name{Masaru \surname{Suzuki}}, 
 \name{Yoshiki \surname{Hidaka}}, 
 and \name{Shoichi \surname{Kai}}}
}

\inst{
Department of Applied Quantum Physics and Nuclear Engineering, 
Kyushu University,
\address{Nishi-ku, Fukuoka 819-0395}
}

\abst{
We theoretically investigate the current--voltage ($I$--$V$) property of two-dimensional Coulomb blockade (CB) arrays by conducting Monte Carlo simulations.
The $I$--$V$ property can be divided into three regions
and we report the dependence of the aspect ratio $\delta$ (namely, the lateral size $N_{y}$ over the longitudinal one $N_{x}$).
We show that the average CB threshold obeys a power-law decay as a function of $\delta$.
Its exponent $\gamma$ corresponds to a sensitivity of the threshold depending on $\delta$,
and is inversely proportional to $N_{x}$ (i.e.,~$\delta$ at fixed $N_{y}$).
Further, the power-law exponent $\zeta$, characterizing the nonlinearity of the $I$--$V$ property in the intermediate region,
logarithmically increases as $\delta$ increases.
Our simulations describe the experimental result $\zeta=2.25$ obtained by Parthasarathy et al.~[Phys.~Rev.~Lett.~{\bf 87} (2001) 186807].
In addition, the asymptotic $I$--$V$ property of one-dimensional arrays obtained by Bascones et al.~[Phys.~Rev.~B.~{\bf 77} (2008) 245422]
is applied to two-dimensional arrays.
The asymptotic equation converges to the Ohm's law at the large voltage limit,
and the combined tunneling-resistance is inversely proportional to $\delta$.
The extended asymptotic equation with the first-order perturbation well describes 
the experimental result obtained by Kurdak et al.~[Phys.~Rev.~B~{\bf 57} (1998) R6842].
Based on our asymptotic equation,
we can estimate physical values that it is hard to obtain experimentally.
}

\kword{Coulomb blockade, nonlinearity, current--voltage property, asymptotic property, power law, size dependence, aspect ratio}

\begin{document}
\maketitle

%%%%%%%
\section{Introduction} \label{sec:introduction}
A Coulomb blockade (CB) \cite{Fulton1987,Heinzel2003} emerges in condensed matter physics, 
and it causes threshold and nonlinear current--voltage ($I$--$V$) behavior.
In some sense, CB can be regarded as a phenomenon that occurs in disordered systems with thresholds 
such as charge-density waves \cite{Gruner1988} and Wigner crystals \cite{Williams1991}.
CB was first studied in a single--electron transistor \cite{Fulton1987, Averin1991}.
Nowadays, it is studied in many systems 
such as arrays of metallic islands \cite{Mooij1990, Rimberg1995, Kurdak1998}, 
metal nanocrystal arrays \cite{Black2000, Parthasarathy2001, Fan2004}, 
molecular arrays \cite{Reed1997, Schoonveld2000, Kane2010, Stokbro2010}, 
a Tomonaga--Luttinger liquid such as a carbon nanotube \cite{Bockrath1999, Yao1999, Monteverde2006},
and graphene quantum-dot arrays \cite{Joung2011}.
Some numerical studies of CB have been done,
e.g.~Monte Carlo (MC) methods \cite{Geigenmuller1989, Middleton1993, Suvakov2007, Bascones2008, Suvakov2010},
molecular dynamics (MD) simulations \cite{Reichhardt2003, Tekic2005, Reichhardt2006},
and circuit dynamics \cite{Oya2007, Kikombo2007}.
In most of these cases, a core work is to investigate its $I$--$V$ property.
It is known that the $I$--$V$ property of typical CB arrays can be divided 
into three characteristic regions according to path flow of electrons \cite{Parthasarathy2001, Reichhardt2003}: 
near the CB threshold $V_{\text{th}}^{(\text{CB})}$, the intermediate voltage region, and the large voltage region.
Several static trajectories exist near the threshold, 
and a crossover from a static to a dynamic trajectory occurs when the bias voltage is set in the intermediate region \cite{Reichhardt2003}.
As the bias voltage increases, trajectories are again static and linear.
The $I$--$V$ property thus approaches to Ohmic behavior in the large voltage region.

The $I$--$V$ property near the CB threshold is mainly characterized by the value of the CB threshold.
The threshold is determined by trajectory of electrons, 
and is thus sensitive to conditions such as the array size and surface disorder.
The $I$--$V$ property is approximately described as
\begin{equation} I \sim (V-V^{(\text{CB})}_{\text{th}})^{\zeta}, \label{MW_power_law} \end{equation}
where $V$ is the bias voltage.
In the intermediate region, the $I$--$V$ property also exhibits nonlinear behavior described as eq.~(\ref{MW_power_law}).
The value of $\zeta$ for several systems has been determined from both experiments and simulations. 
For example, an experimental study shows that an array of normal metal islands has $\zeta = 1.36~\pm~0.1$ for a one-dimensional (1D) array 
and $\zeta = 1.8~\pm~0.16$ for a two-dimensional (2D) square array \cite{Rimberg1995};
in addition, other experimental studies show that a metal nanocrystal has $\zeta = 2.25~\pm~0.1$ for a 2D triangle array \cite{Parthasarathy2001},
and that a gold nanocrystal has $\zeta = 2.7$ to $3.0$ for a 3D array \cite{Fan2004}.
Experiments of colloidal deposition show $\zeta = 2.1$ for 2D and $\zeta = 3.5$ for 3D \cite{Lebreton1998}.
For numerical simulations, MC calculations show that $\zeta = 1.0$ for linear arrays 
and $\zeta = 2.0$ for square arrays \cite{Middleton1993}, 
and MD calculations show that $\zeta = 1.94~\pm~0.15$ for square arrays \cite{Reichhardt2003}.
In a theoretical study, a mean-field analysis suggested that $\zeta = 2$ for a 2D array \cite{Roux1987}.
By analyzing surface evolution on arrays with charge disorder, $\zeta = 5/3$ is analytically predicted \cite{Middleton1993}. 
Here, we should emphasize that $\zeta$ has been discussed in relation to the array configuration and array dimension so far;
the size dependence has not been taken into account.
Meanwhile, some previous studies \cite{Lebreton1998, Black2000, Reichhardt2003} mention 
that the exponent $\zeta$ monotonically increases with increasing the lateral size from a 1D linear array to a 2D square array.
However, they have qualitatively focused on only the arrays in which the lateral size is less than the longitudinal one
and expected the exponent to be constant for large lateral size.

% configurations
\begin{figure*}[!t]
\begin{center}
\includegraphics[width=160mm]{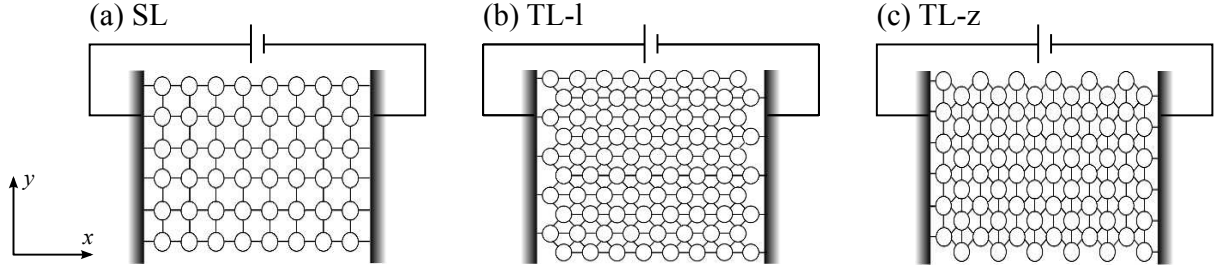}
\end{center}
\caption{
The present configurations: (a) a simple lattice (SL), (b) a line-type triangular lattice (TL-l), and (c) a zigzag-type triangular lattice (TL-z).
Each circle indicates a Coulomb island.
This SL contains $N=48$ islands ($N_{x}=8$, $N_{y}=6$, and $\delta=0.75$),
this TL-l does $N=80$ islands ($N_{x}=8$, $N_{y}=10$, and $\delta=1.25$),
and this TL-z does $N=72$ islands ($N_{x}=12$, $N_{y}=6$ and $\delta=0.5$).
The islands are sandwiched between the positive (left) and negative (right) electrodes, 
and each solid line between island-island or island-electrode represents the tunneling junction.
Each island touches the gate electrode (not displayed in the figure). 
}
\label{fig:conf_layout}
\end{figure*}

Middleton and Wingreen (MW) explicitly introduced offset charge distribution in their model.
The charge disorder originates from the surface impurity.
In their model, Bascones et al.~have discussed the asymptotic $I$--$V$ property 
of 1D arrays in the large voltage region \cite{Bascones2008}.
It converges to the Ohm's law at the large voltage limit.
In addition, they showed the presence of the offset voltage $V_{\text{offset}}$,
and analytically expressed it in short-limit of the interaction range.

In this paper, we carry out MC simulations to study the size dependence of the $I$--$V$ property 
for configurations such as a simple lattice and a triangular lattice.
We employ the model proposed by MW \cite{Middleton1993}.
Based on their model, we extend it to size dependence.
Our main results are the following:
(i)~the average CB threshold $\bar{V}_{\text{th}}^{(\text{CB})}$,
(ii)~the power-law exponent $\zeta$ in the intermediate voltage region, 
and (iii)~an asymptotic $I$--$V$ curve of 2D arrays with the first-order perturbation of $\varepsilon$.

This paper is organized as follows. 
In \S~\ref{sec:simulation}, we briefly describe the present configurations, simulation model, and numerical conditions. 
In \S~\ref{sec:RandD},  we first express the size dependence of the average CB threshold for simple configurations,
and then, the size dependences of the exponent $\zeta$ is shown for several configurations.
In addition, we express the asymptotic $I$--$V$ property obtained analytically for simple configurations, 
and then we compare it with simulation and experimental results to verify the asymptotic relation.
In \S~\ref{sec:summary}, we summarize our results.

%%%%%%
\section{Method} \label{sec:simulation}
\subsection{Structure}
The simplest single-electron transistor consists of two tunnel junctions that connect the source and drain electrode, respectively,  
and the region sandwiched between the tunnel junctions also connects to the gate electrode through the gate capacitor.
The sandwiched region, known as the Coulomb island, can be regarded as a place where charge accumulates.
We consider arrays constructed of Coulomb islands between positive and negative electrodes. 
A series of tunneling processes cause electrons to flow in the arrays. 
Each island also connects to the gate electrode with the gate capacitor.
The i-th island has charge $Q_{\text{i}}$ and potential $\Phi_{\text{i}}$.
The charge $Q_{\text{i}}$ contains both an integer multiple of the elementally charge $ne$ (where $n$ denotes an integer and $e$ the elementary charge)
and offset charge $-e/2 \le q_{\text{i}} \le e/2$ due to the impurity \cite{Middleton1993}. 
In simulations, the offset charges are set by uniform random numbers and remain constant over time.

Three configurations are considered (Fig.~\ref{fig:conf_layout}).
The first configuration is a simple lattice (SL), 
and the remaining two configurations involve different directions of a triangular lattice: 
a line-type triangular lattice (TL-l) and a zigzag-type triangular lattice (TL-z).
We set $x$- and $y$-directions as shown in Fig.~\ref{fig:conf_layout}, 
namely $x$-direction corresponds to longitudinal direction and $y$-direction does to lateral one. 
The SL configuration is characterized by the number of horizontal islands $N_{x}$ and vertical islands $N_{y}$. 
Thus, the total number of islands is $N=N_{x}\times N_{y}$. 
Both the TL-l and TL-z configurations are also characterized by $N_{x}$ and $N_{y}$, 
and the total number of islands is $N=N_{x}\times N_{y}$. 
Although we use $N_{x}$ and $N_{y}$ in every configurations, 
the method of setting them is underspecified for triangular lattices.
Here, we define $N_{x}$ and $N_{y}$ as described in the caption of Fig.~\ref{fig:conf_layout},
and the aspect ratio $\delta$ is defined as the lateral size over the longitudinal size; i.e., $\delta=N_{y}/N_{x}$.

To calculate the total energy of the array, it is useful to consider a configuration matrix.
The physical configuration of the lattices uniquely determines the configuration matrix whose element $M_{\text{ij}}$ is represented as
\begin{equation} M_{\text{ij}}=\delta_{\text{ij}}\left[\sum_{\text{k}}C_{\text{ik}}+\sum_{\mu = +, -, g}C_{\text{i},\mu}\right]-C_{\text{ij}} \label{def_capacitance_matrix} \end{equation}
where $C_{\text{ij}}$ denotes the tunneling capacitance between the i-th and j-th islands, 
$C_{\text{i},\mu}$ the tunneling capacitance between the i-th island and the electrode $\mu\in \{+, -, g\}$, 
and $\delta_{\text{ij}}$ the Kronecker delta. 
Note that the symbols $+$, $-$, and $g$ indicate the positive, negative, and gate electrodes, respectively.
If the i-th island does not connect to the j-th one, then $C_{\text{ij}}$ is set to 0.
Electrons move through the network of islands, while the islands themselves do not move.
Therefore, the configuration matrix $M_{\text{ij}}$ remains constant over time in this study. 

\subsection{Model}
We briefly summarize the time-evolution procedure used in the MC method \cite{Geigenmuller1989, Middleton1993} in this subsection.

The system evolves to decrease the total electrostatic energy $E$, whose derivation is summarized in Appendix \ref{appendix:total_energy}.
In MC simulations, the electrons are virtually moved for each possible tunneling event. 
We can calculate the energy change $\Delta E_{\text{n}^{\prime}\to \text{m}^{\prime}}$ at $\text{n}^{\prime} \to \text{m}^{\prime}$ (see Appendix \ref{appendix:energy_change}), where $\{\text{n}^{\prime},\ \text{m}^{\prime}\}\in \{1,\ 2,\ \dots\ , N,\ +,\ -\}$. 
The tunneling rate $\Gamma_{\text{n}^{\prime}\to \text{m}^{\prime}}$ at $\text{n}^{\prime}\to\text{m}^{\prime}$ is calculated as \cite{Likharev1986}
\begin{eqnarray} \Gamma_{\text{n}^{\prime} \to \text{m}^{\prime}} = \frac{1}{e^{2}R_{\text{t},\text{n}^{\prime}\to \text{m}^{\prime}}} \frac{-\Delta E_{\text{n}^{\prime}\to \text{m}^{\prime}}}{1-\exp\left[\Delta E_{\text{n}^{\prime}\to \text{m}^{\prime}}/k_{B}T\right]}, \end{eqnarray} 
where $R_{\text{t},\text{n}^{\prime}\to \text{m}^{\prime}}$ denotes the tunneling resistance at $\text{n}^{\prime}\to\text{m}^{\prime}$.
It is assumed that the tunnel resistance between an island and the gate electrode is infinity in these calculations, 
i.e.,~the electrons cannot move between an island and the gate electrode. 
The tunneling rate is derived by assuming that the tunneling events occur independently.
The resistances $R_{\text{t},\text{n}^{\prime}\to \text{m}^{\prime}}$ depend on the configuration of the array in general, 
but we regard them to be a constant $R_{\text{t}}$, where $R_{\text{t},\text{n}^{\prime}\to \text{m}^{\prime}}=R_{\text{t}}$. 

Assuming Poisson distribution, the probability distribution that a tunneling from $\text{n}^{\prime}$ to $\text{m}^{\prime}$ occurs {\it at} time lag $t$ is represented as
\begin{equation} f_{\text{n}^{\prime}\to \text{m}^{\prime}}(t)=\Gamma_{\text{n}^{\prime}\to \text{m}^{\prime}}\exp\left[-\Gamma_{\text{n}^{\prime}\to \text{m}^{\prime}}t\right]. \end{equation}
Then, the cumulative distribution is equivalent to the distribution that a tunneling from $\text{n}^{\prime}$ to $\text{m}^{\prime}$ occurs {\it during} time lag $t-t_{0}$, represented as
\begin{equation} F_{\text{n}^{\prime}\to \text{m}^{\prime}}(t)=1-\exp\left[-\int_{t_{0}}^{t}\Gamma_{\text{n}^{\prime}\to \text{m}^{\prime}}(t^{\prime}){\rm d}t^{\prime}\right], \end{equation}
where $t_{0}$ denotes the time when the last tunneling event from $\text{n}^{\prime}$ to $\text{m}^{\prime}$ occurred. 
The energy changes depend on time, and hence the tunneling rate also does.
A uniform random number $x=x(t)$ in $[0,1]$ is introduced and the cumulative distribution is set as $F_{\text{n}^{\prime}\to \text{m}^{\prime}}(t)=x$.
Note that the random number $x$ is updated only when a tunneling event from $\text{n}^{\prime}$ to $\text{m}^{\prime}$ occurs. 
We thus obtain the time interval between the last tunneling event from $\text{n}^{\prime}$ to $\text{m}^{\prime}$ and the next one as
\begin{equation} \Delta t_{\text{n}^{\prime}\to \text{m}^{\prime}}=\frac{-\log(1-x)-\sum_{\text{k}=0}^{K-1}\Gamma_{\text{n}^{\prime}\to \text{m}^{\prime}}(t_{\text{k}})\Delta t_{\text{k}}}{\Gamma_{\text{n}^{\prime}\to \text{m}^{\prime}}(t)}, \end{equation}
where $K$ denotes the number of tunneling events within the entire array during $t-t_{0}$. 
Note that $\Gamma_{\text{n}^{\prime}\to \text{m}^{\prime}}$ remains constant over time $\Delta t_{\text{k}}$, 
and $\Delta t_{\text{n}^{\prime}\to \text{m}^{\prime}}$ is not equal to $t-t_{0}$ in general. 
We use the smallest $\Delta t_{\text{n}^{\prime}\to \text{m}^{\prime}}$ for the time evolution increments.

\subsection{Simulation}
Below the CB threshold $V_{\text{th}}^{(\text{CB})}$, the tunneling interval $\Delta t$ is infinity for each path.
Therefore, we can determine the threshold voltage above which $\Delta t$ is finite in the steady state.
The current along the path $\text{n}^{\prime}\to \text{m}^{\prime}$ can be calculated as
\begin{equation} I_{\text{n}^{\prime}\to \text{m}^{\prime}}=-e\left(\Gamma_{\text{n}^{\prime}\to \text{m}^{\prime}}-\Gamma_{\text{m}^{\prime}\to \text{n}^{\prime}}\right). \label{current_def} \end{equation}
In simulations, the current can be calculated along all path, 
but we focus only on those paths that neighbor the positive or negative electrodes, represented as
\begin{equation} I_{\text{posi}} := {\sum_{\text{i}}}^{\prime}I_{+\to \text{i}}\ ,\ \ I_{\text{nega}} := {\sum_{\text{i}}}^{\prime}I_{\text{i}\to -}\ , \end{equation}
where the sigma with the prime denotes summation over only those islands that neighbor the positive or negative electrodes.
In the steady state, $I_{\text{posi}}$ and $I_{\text{nega}}$ are the same because of Kirchhoff's current law.
Hence, we demonstrate only $I_{\text{posi}}$ as the current $I$ in the remaining sections.

The voltages of the negative and gate electrodes are fixed at $\Phi_{-}=\Phi_{g}=0$, 
and the voltage of the positive electrode $\Phi_{+}$ is thus adjusted as the control parameter, 
i.e.,~the bias voltage $V$ is equivalent to the voltage of the positive electrode $\Phi_{+}$. 
The initial condition was $Q_{\text{i}}=q_{\text{i}}$ and $\Phi_{+}=0$. 
The voltage $\Phi_{+}$ was incremented by $\Delta \Phi_{+}$,
and before we sampled the physical variables at each voltage, we waited sufficiently long for the system to return to the steady state. 
Note that this waiting time depends on system conditions, such as the configuration and $\Delta \Phi_{+}$.

We assume that the system is at zero temperature.
The capacitance is set at $C=10^{-4}C_{g}$, 
Note that the ratio $\varepsilon:=C/C_{g}$ corresponds to the interaction range\cite{Middleton1993}.
The increment voltage $\Delta \Phi_{+}$ is $10^{-2}$;
therefore the threshold voltage has an uncertainty of the order $10^{-3}$.
Finally, the charge is scaled by $e$, 
the capacitance by $C_{g}$, 
the time by $R_{\text{t}}C_{g}$, 
the current by $e/R_{\text{t}}C_{g}$, 
the potential by $e/C_{g}$, 
and the energy by $e^{2}/C_{g}$. 

%%%%%%%
\section{Results and Discussion} \label{sec:RandD}
\subsection{Size dependences of the average threshold}
% Vth vs width (log)
\begin{figure}[!t]
\begin{center}
\includegraphics[width=85mm]{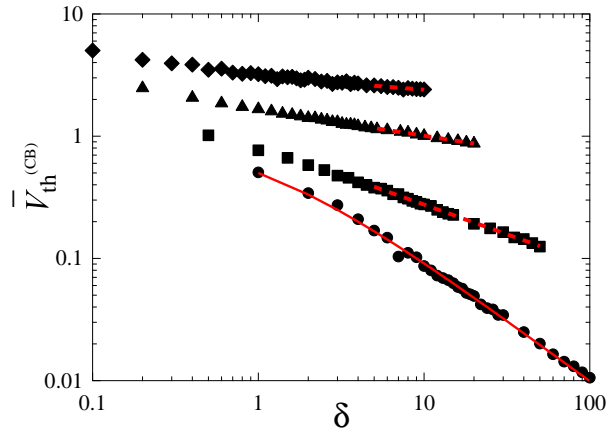}
\end{center}
\caption{(Color online)
Log--log plot of average CB threshold $\bar{V}_{\text{th}}^{(\text{CB})}$ versus the aspect ratio $\delta$
for several longitudinal sizes; $N_{x}=1$ (circle), $2$ (square), $5$ (triangle), and $10$ (diamond).
The red solid line represents eq.~(\ref{Vth_width_Nh001}), 
and the red dashed line represents power-law fitting.
The data in $\delta \ge 5$ are used to obtain the fitting parameters.
}
\label{fig:Vth_vs_width_log}
\end{figure}

% exp_vs_Nh
\begin{figure}[!t]
\begin{center}
\includegraphics[width=72mm]{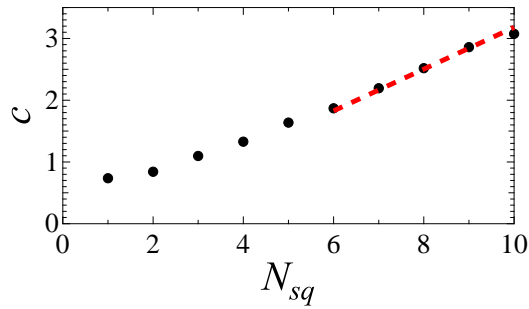}
\end{center}
\caption{(Color online) 
The coefficient $c$ of eq.~(\ref{CBVth_power_law}) versus a length of a side of square arrays.
A filled circle represents a simulation result.
The red dashed line is a line with slope 0.338.
}
\label{fig:c_vs_Nh}
\end{figure}

% exp_vs_Nh
\begin{figure}[!t]
\begin{center}
\includegraphics[width=72mm]{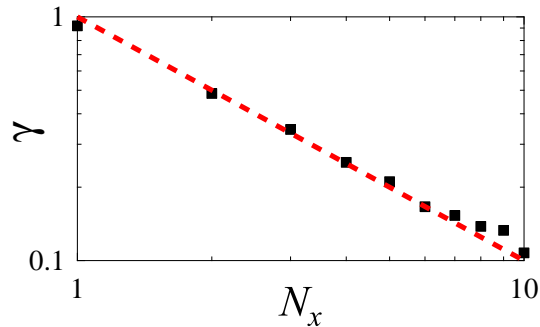}
\end{center}
\caption{(Color online)
The exponent $\gamma$ of eq.~(\ref{CBVth_power_law}) as a function of $N_{x}$ (i.e., $\delta$ at fixed $N_{y}$).
A filled square represents a simulation result.
The dashed line represents eq.~(\ref{gamma_power_law}).}
\label{fig:exp_vs_Nh}
\end{figure}

We first investigate the size dependences of the average CB threshold for SL.
Figure \ref{fig:Vth_vs_width_log} shows the average CB threshold $\bar{V}_{\text{th}}^{(\text{CB})}$ 
as a function of the aspect ratio $\delta$ for several longitudinal size $N_{x}$. 
Each point on the figure is derived from the average of at least 50 different initial distributions of the offset charges. 
Our results of 1D arrays (i.e., $N_{y}=1$) are in agreement with the previous results by MW \cite{Middleton1993}.

The average threshold is analytically represented as
\begin{equation} \bar{V}^{(\text{CB})}_{\text{th}}(N_{x}, N_{y})=\int_{-e/2}^{e/2}\frac{{\rm d}q_{1}}{e}\cdots\int_{-e/2}^{e/2}\frac{{\rm d}q_{N}}{e} V_{\text{th}}^{(\text{CB})}(\{q_{\text{i}}\}). \end{equation}
For the simplest case $N_{x}=N_{y}=1$ and $\varepsilon \ll 1$, 
the CB threshold voltage as a function of the initial charge $q$ is obtained as
\begin{equation} V_{\text{th}}^{(\text{CB})}(q)= \frac{e}{C_{g}}\left(\frac{q}{e}+\frac{1}{2}\right), \end{equation}
namely, $V_{\text{th}}^{(\text{CB})}$ of a single Coulomb island is proportional to the initial offset charge.
For $N_{x}=1$ (i.e., $N=N_{y}$) and $\varepsilon \ll 1$, 
the threshold is dominated by the smallest initial offset charge.
The average threshold reduces to
\begin{eqnarray} & &  \bar{V}_{\text{th}}^{(\text{CB})}(N_{x}=1, N_{y}) \cr
& & \cr
& = & \frac{e}{C_{g}} N_{y}! \int_{-1/2}^{1/2}{\rm d}q^{\prime}_{1}\int_{-1/2}^{q^{\prime}_{1}}{\rm d}q^{\prime}_{2} \cdots \int_{-1/2}^{q^{\prime}_{N-1}}{\rm d}q^{\prime}_{N}\left(q^{\prime}_{N}+\frac{1}{2}\right), \cr
& &  \end{eqnarray}
where $\{q^{\prime}_{\text{i}}\}$ denote the reordered dimensionless charges: $1/2>q^{\prime}_{1} > q^{\prime}_{2} > \cdots > q^{\prime}_{N}>-1/2$.
The average threshold for $N_{x}=1$ is thus obtained as
\begin{equation} \bar{V}_{\text{th}}^{(\text{CB})}(N_{x}=1, N_{y})=\frac{e}{C_{g}}\frac{1}{N_{y}+1}, \label{Vth_width_Nh001} \end{equation}
and this well describes the simulation result as shown in Fig.~\ref{fig:Vth_vs_width_log}.
For $N_{x}>1$, it is difficult to derive the average threshold 
because electron meandering plays an important role just above $V_{\text{th}}^{(\text{CB})}$.
Nevertheless, we find that the average CB threshold for large $\delta$ can be described by a power law
\begin{equation} \bar{V}^{(\text{CB})}_{\text{th}}(N_{x}, N_{y}) = c\delta^{-\gamma}. \label{CBVth_power_law} \end{equation}
In fact, eq.~(\ref{Vth_width_Nh001}) for large $N_{y}$ implies the power-law decay as eq.~(\ref{CBVth_power_law}).

The proportionality coefficient $c$ of eq.~(\ref{CBVth_power_law}) indicates the value of the average threshold for square arrays (i.e.,~$N_{x}=N_{y}$).
Figure \ref{fig:c_vs_Nh} shows the coefficient $c$ as a function of $N_{sq}$ that is a length of a side of square arrays. 
MW have reported $\bar{V}_{\text{th}}^{(\text{CB})}\sim 0.338 N_{sq}$ for square arrays \cite{Middleton1993},
and the line is plotted in Fig.~\ref{fig:c_vs_Nh}.
The simulation results deviate from the line in small $N_{sq}$ region.
This is because we cannot regard the array with $N_{x}=N_{y}=1$ as a square array,
namely, $N_{x}$ is too small to regard arrays as square in that region.
In addition, the simulation results do not satisfy the power law near $\delta \simeq 1$ at small $N_{x}$.

The power-law exponent $\gamma$ of eq.~(\ref{CBVth_power_law}) is shown in Fig.~\ref{fig:exp_vs_Nh}.
The relation $\gamma \simeq 1$ for $N_{x}=1$ is evident from eq.~(\ref{Vth_width_Nh001}).
The exponent $\gamma$ is thus interpreted as a sensitivity of the threshold depending on $\delta$ comparing to arrays with $N_{x}=1$.
As $N_{x}$ increases, the increment of the aspect ratio decreases even for the same increment of $N_{y}$.
Therefore, it can be expected that $\gamma$ is a monotonically decreasing function of $N_{x}$.
In fact, as shown in Fig.~\ref{fig:exp_vs_Nh}, the exponent $\gamma$ is inversely proportional to $N_{x}$, namely,
\begin{equation} \gamma(N_{x}) = {N_{x}}^{-1}. \label{gamma_power_law} \end{equation}
These results will be a hint to understand the size dependences of the CB threshold.
In addition, it is interesting to observe the power law decay in experiments.

\subsection{Logarithmic increase of the power-law exponent $\zeta$}
% IVplot for several Nh
\begin{figure}[!t]
\begin{center}
\includegraphics[width=75mm]{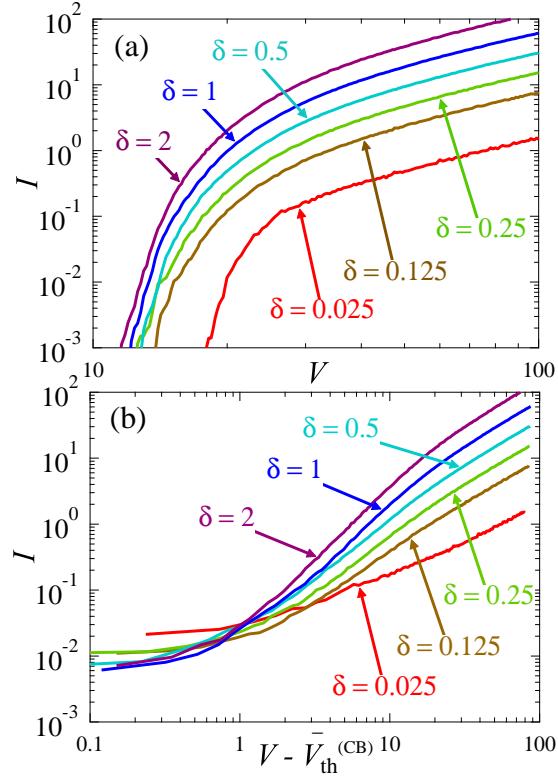}
\end{center}
\caption{(Color online)
Average current is shown for SL with $N_{x}=40$.
(a) $I$--$V$ plot for the aspect ratio $\delta = 0.025,~0.125,~ 0.25,~0.5,~1,~\text{and}~2$. 
(b) Current as a function of $V-\bar{V}_{\text{th}}^{(\text{CB})}$ for the same values of $\delta$.}
\label{fig:I_V_diffw}
\end{figure}

\begin{figure}[!t]
\begin{center}
\includegraphics[width=82mm]{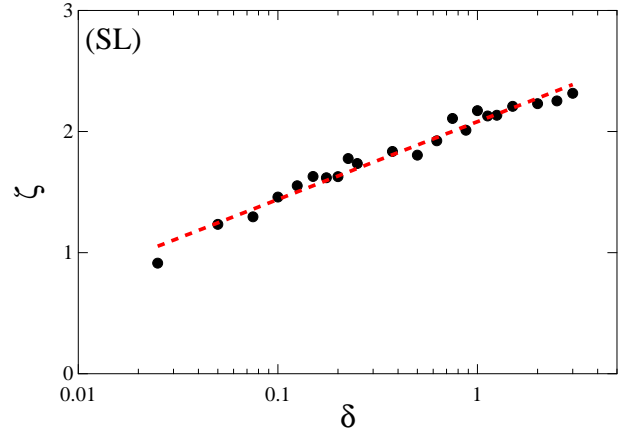}
\end{center}
\caption{(Color online)
Semi-log plot of the exponent $\zeta$ for the simple lattice as a function of the aspect ratio $\delta$ with $N_{x}=40$.
This plot is extracted from the average $I$--$V$ relation shown in Fig.~\ref{fig:I_V_diffw} (b).
The red dashed line shows the fitting line determined by eqs.~(\ref{logarithmic}) and (\ref{zeta_line}) with $\zeta^{(\text{sq})}=2.08$ and $\zeta^{(\text{line})}=1.05$.
}
\label{fig:zeta_vs_width}
\end{figure}

We next investigate the exponent $\zeta$ in eq.~(\ref{MW_power_law}) as a function of the aspect ratio $\delta$.
Figure \ref{fig:I_V_diffw} shows the averaged $I$--$V$ property for several $\delta$, 
and each curve results from the average of at least 30 data sets.
The nonlinear behavior is more visible in Fig.~\ref{fig:I_V_diffw} (b).
Further, Fig.~\ref{fig:zeta_vs_width} shows the exponent $\zeta$ as a function of $\delta$ with $N_{x}=40$.
The exponents are obtained by fitting to the average $I$--$V$ property in the intermediate region defined as $10^{0.5} < V-\bar{V}_{\text{th}}^{(\text{CB})} < 10$.
The fitting range is selected to prevent artificiality from being included into the value of the exponents.
The aspect ratio dependence of $\zeta$ appears to be approximately represented by
\begin{equation} \zeta = \zeta^{(\text{sq})} + b\log_{10}\delta \label{logarithmic} \end{equation}
with fitting parameters $\zeta^{(\text{sq})}$ and $b=b(N_{x})$.
The $\zeta^{(\text{sq})}$ parameter denotes the exponent of the square array, 
and $\zeta^{(\text{sq})}~\simeq~2.08$ approximately agrees with the previous result \cite{Middleton1993}.
Using the $b$ parameter, the exponent $\zeta^{(\text{line})}$ for a 1D simple array, i.e.,~$N_{y}=1$, is represented as
\begin{equation} \zeta^{(\text{line})}=\zeta^{(\text{sq})}-b\log_{10}N_{x}, \label{zeta_line} \end{equation}
and $\zeta^{(\text{line})}~\simeq~1.05$ is obtained as shown in Fig.~\ref{fig:zeta_vs_width}.
MW have reported that, for arbitrary $N_{x}$, the exponents of linear and square arrays are $\zeta\simeq 1$ and $\zeta \simeq 2$, respectively \cite{Middleton1993}.
The same logarithmic increase is thus expected to appear in different longitudinal size (i.e.,~$N_{x}$),
while we only show single longitudinal size (i.e., $N_{x}=40$) in Fig.~\ref{fig:zeta_vs_width}.
Moreover, although our simulations are only performed for finite $\delta$ in each configuration,
it is expected that the exponent diverges logarithmically at such large values of $\delta$.

% width dependence of zeta

% Triangle Lattice (Tr)
\begin{figure}[!h]
\begin{center}
\includegraphics[width=80mm]{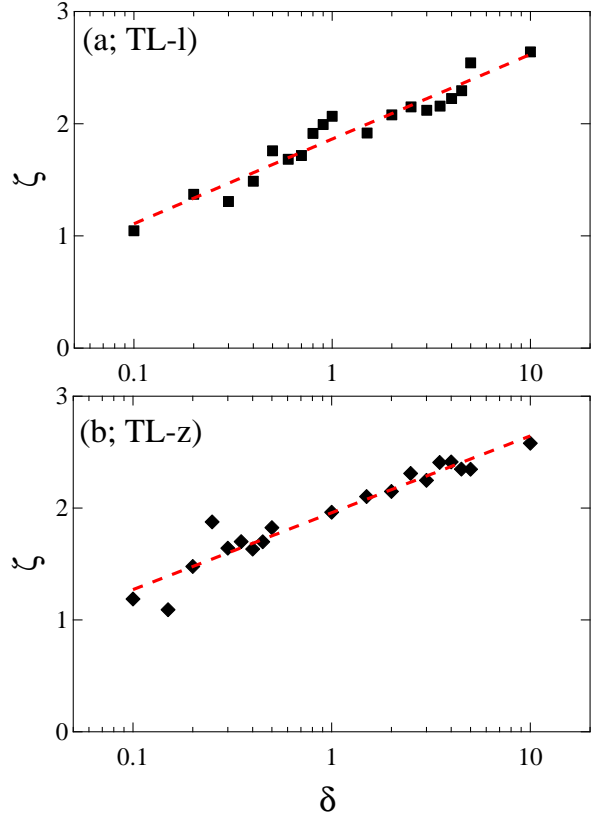}
\end{center}
\caption{(Color online)
Semi-log plot of the exponent $\zeta$ for (a) TL-l and (b) TL-z as a function of the aspect ratio $\delta$ with $N_{x}=20$.
The red dashed lines show the fitting lines determined by eqs.~(\ref{logarithmic}) and (\ref{zeta_line}) 
with $\zeta^{(\text{sq})}=1.86$ and $\zeta^{(\text{line})}=0.884$ for TL-l and with $\zeta^{(\text{sq})}=1.96$ and $\zeta^{(\text{line})}=1.06$ for TL-z.
}
\label{fig:zeta_vs_width_Tr_layout}
\end{figure}

We also focus on the triangular lattices.
Figures \ref{fig:zeta_vs_width_Tr_layout} (a) and (b) show the exponent $\zeta$ as a function of the aspect ratio $\delta$ for TL-l and TL-z, respectively.
Note that, similar to SL, the $I$--$V$ properties are averaged over at least 30 data sets 
and the exponent is extracted from the averaged results.
Here, the intermediate region is defined as $10^{0.4} < V-\bar{V}_{\text{th}}^{(\text{CB})} < 10^{0.6}$, and the fitting is done in that range.
Similar to the exponent of SL, the exponents $\zeta$ of TL-l and TL-z increase logarithmically as determined by eqs.~(\ref{logarithmic}) and (\ref{zeta_line}).
Parthasarathy et al.~\cite{Parthasarathy2001} experimentally showed $\zeta~\simeq~2.25$ for a well-ordered triangular array of gold nanocrystals 
with an array size of $N_{x}=30$ to $90$ and $N_{y}\simeq 270$.
In the range $3 \le \delta \le 9$, our simulation results of both TL-l and TL-z show $\zeta\simeq 2.25$.
Our results propose that we should pay attention to the aspect ratio as well as the array configuration and array dimension
when discussing the exponent $\zeta$.

We should recall that the universality of the logarithmic increase requires careful attention.
The above results are obtained in locally coupled CB, i.e.,~small $\varepsilon$.
The behavior of $\zeta$ for large $\varepsilon$ may differ from that for small $\varepsilon$, 
because the interaction among electrons ranges over the entire array in large $\varepsilon$ systems.
The $\varepsilon$ dependences are still an open question.

\subsection{Analytical asymptotic equation at large bias voltages}
We finally discuss the $I$--$V$ property for large values of the bias voltage $V=\Phi_{+}-\Phi_{-}$.
As mentioned above, Bascones et al.~have derived the asymptotic $I$--$V$ property 
of 1D arrays with the offset voltage $V_{\text{offset}}$
Their offset voltage $V_{\text{offset}}$ is expressed in the limit of $\varepsilon\to 0$ \cite{Bascones2008}.
We extend their study to 2D simple configurations.
In addition, our extended expression of $V_{\text{offset}}$ contains the first-order perturbation of $\varepsilon$.

\begin{figure}[!t]
\begin{center}
\includegraphics[width=84mm]{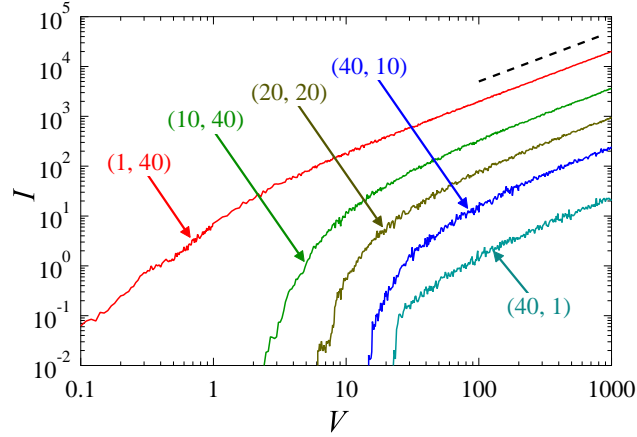}
\end{center}
\caption{(Color online)
Current--voltage plot for several sizes of SL at $T=0$, 
where $(N_{x},~N_{y}; ~\delta)$=$(1,~40;~40)$, $(10,~40;~4)$, $(20,~20;~1)$, 
$(40,~10;~0.25)$, and $(40,~1;~0.025)$ from left to right.
The dashed line represents linear behavior.
}
\label{fig:IVrelation}
\end{figure}

Figure \ref{fig:IVrelation} shows the $I$--$V$ plot for several sizes of the SL configuration.
All results exhibit linear behavior at large bias voltage limit, and the asymptotic curve can be obtained as follows.
The energy changes of the 1D array that contains $N_{x}$ islands reduce to (see Appendix \ref{appendix:energy_change})
\begin{subequations}
\begin{eqnarray} \Delta E_{1\to +} & = & e\left(V_{1}+V_{1}^{(\text{ext})}\right)-e\Phi_{+}+\frac{e^{2}}{2}M_{11}^{-1},~~~ \\
\Delta E_{\text{n}+1 \to \text{n}} & = & e\left(V_{\text{n}+1} +V_{\text{n}+1}^{(\text{ext})}-V_{\text{n}}-V_{\text{n}}^{(\text{ext})}\right) \cr
& & +\frac{e^{2}}{2}\left(M_{\text{nn}}^{-1}+M_{\text{n}+1,\text{n}+1}^{-1}-2M_{\text{n},\text{n}+1}^{-1}\right),~~~~ \\
\Delta E_{- \to N_{x}} & = & -e\left(V_{N_{x}}+V_{N_{x}}^{(\text{ext})}\right)+e\Phi_{-}+\frac{e^{2}}{2}M_{N_{x}N_{x}}^{-1},~~~~~~~~ \end{eqnarray}
\end{subequations}
with n $= 1, 2, \dots, N_{x}-1$.
At large bias voltages, the current from the negative to the positive electrode is neglected.
Thus, the all energy changes should be always the same in the large voltage region.
We can obtain $\Delta E:=\Delta E_{1\to +} = \Delta E_{2\to 1}=\cdots = \Delta E_{- \to N_{x}}$ as
\begin{equation} \Delta E = -\frac{e}{N_{x}+1}\left(V - V_{\text{offset}}\right). \end{equation}
Here, the above representation contains an offset voltage $V_{\text{offset}}$ defined as
\begin{equation} V_{\text{offset}}:= e\left(\sum_{\text{i}=1}^{N_{x}} M^{-1}_{\text{ii}}-\sum_{\text{i}=1}^{N_{x}-1}M^{-1}_{\text{i},\text{i}+1} \right). \end{equation}
The offset voltage $V_{\text{offset}}$ differs from the CB threshold $V_{\text{th}}^{(\text{CB})}$. 
For $\varepsilon \ll 1$, 
$V_{\text{offset}}$ can be analytically derived within the first-order perturbation of $\varepsilon$ (see Appendix \ref{appendix:phi_th}) as
\begin{equation}V_{\text{offset}} \simeq \frac{eN_{x}}{C_{g}}\left(1-\frac{3N_{x}-1}{N_{x}}\varepsilon\right). \label{phi_th_epsilon} \end{equation}
The energy changes cannot be the same below the offset voltage $V_{\text{offset}}$.
In contrast, above the offset voltage $V_{\text{offset}}$, 
electrons can move from the negative to positive electrode for any offset charge distributions, 
i.e.,~$V_{\text{offset}}$ is the maximum of $V_{\text{th}}^{(\text{CB})}$ when $\varepsilon \ll 1$ \cite{Narumi_DistVth_2011}.
However, near $V_{\text{offset}}$,
the influence of the offset charge distribution cannot be neglected
because the charge of electrons is discrete.
With increasing the bias voltage, 
the energy changes become sufficiently large to neglect this discreteness.
Therefore, all energy changes in the large voltage region can be always regarded as the same.

% IVrelation

Using eq.~(\ref{current_def}), the asymptotic current is obtained as
\begin{equation}I^{(\text{asy, 1D})}= \frac{V - V_{\text{offset}}}{(N_{x}+1)R_{\text{t}}} \left[1-\exp \left\{-\frac{e\left(V-V_{\text{offset}}\right)}{(N_{x}+1)k_{B}T}\right\}\right]^{-1}. \label{IVrelation_asy} \end{equation}
Note that it is independent of the gate voltage.
The simple array can be easily extended to higher-dimensional arrays.
For the 2D array in which the number of islands is $N_{x}\times N_{y}$, 
the lateral current (i.e.,~the $y$-direction current in Fig.~\ref{fig:conf_layout}) can be neglected at large bias voltages.
This 2D array can be assumed to be composed of the isolated $N_{y}$ 1D arrays that consist of $N_{x}$ islands.
Hence, we obtain
\begin{eqnarray} & & I^{(\text{asy, 2D})} \cr
& = & \frac{N_{y}\left(V-V_{\text{offset}}\right)}{(N_{x}+1)R_{\text{t}}} \left[1-\exp \left\{-\frac{e\left(V-V_{\text{offset}}\right)}{(N_{x}+1)k_{B}T}\right\}\right]^{-1}. \cr
& & \label{IVrelation_asy_2D} \end{eqnarray}
The above equation does not hold when $k_{B}T \to \infty$, 
because the assumption that the current from the negative to positive electrode is negligible is no longer correct at these high temperatures.
In contrast, in finite temperature,
eq.~(\ref{IVrelation_asy_2D}) represents that the asymptotic equation converges in the limit of $V/V_{\text{offset}}\to \infty$
to the Ohmic behavior $V\simeq R_{\text{c}}I$ with the combined tunneling-resistance 
\begin{equation} R_{\text{c}}:=\frac{N_{x}+1}{N_{y}}R_{\text{t}}. \end{equation}
Namely, the combined tunneling-resistance is inversely proportional to the aspect ratio $\delta$; 
$R_{\text{c}}/R_{\text{t}} \simeq \delta^{-1}$ at large $N_{x}$.

% IVrelation_asy
\begin{figure}[!t]
\begin{center}
\includegraphics[width=85mm]{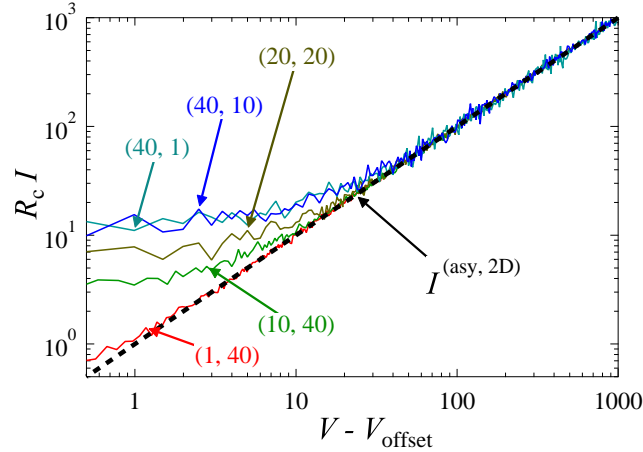}
\end{center}
\caption{(Color online)
Current--voltage plot for several sizes $(N_{x}, N_{y})$ of the SL configuration at $T=0$ for the results shown in Fig.~\ref{fig:IVrelation};
the asymptotic line calculated from eq.~(\ref{IVrelation_asy_2D}) is shown by the black dashed line.
The horizontal axis denotes $V-V_{\text{offset}}$ and the vertical axis denotes $R_{\text{c}}I$ with the combined resistance $R_{\text{c}}=(N_{x}+1)R_{\text{t}}/N_{y}$.
All results for large $V$ collapse to the asymptotic line. 
}
\label{fig:IVrelation_asy}
\end{figure}

As shown in Fig.~\ref{fig:IVrelation_asy}, which plots $R_{\text{c}}I$ as function of  $V-V_{\text{offset}}$, 
the asymptotic line calculated from eq.~(\ref{IVrelation_asy_2D}) completely describes simulation results 
of arbitrary $\delta$ at large bias voltages (roughly, $V/V_{\text{offset}}>2$).
Near the offset voltage $V_{\text{offset}}$, the simulation results deviate from the asymptotic line
because each energy change is different from the others.
As mentioned above, this originates from discreteness of charge.
In fact, as $N_{x}$ decreases (i.e.,~the number of the energy changes decreases),
the results collapse to the asymptotic line at smaller $V-V_{\text{offset}}$.

\begin{figure}[!t]
\begin{center}
\includegraphics[width=85mm]{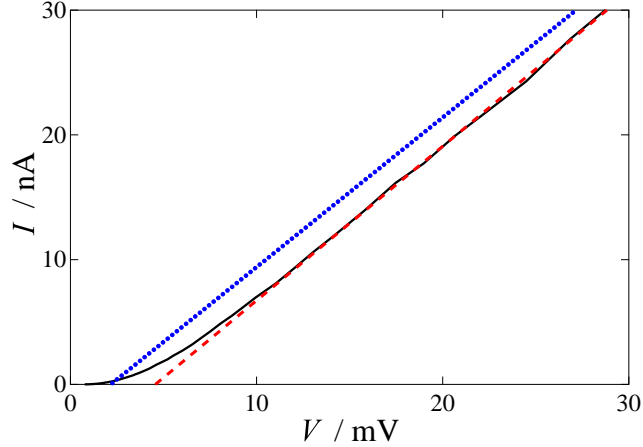}
\end{center}
\caption{(Color online)
Current--voltage property for the experimental result extracted from figure 2 (sample A) in reference \cite{Kurdak1998} (black solid curve)
and the analytical result calculated from the following formula and values given in reference \cite{Kurdak1998} (blue dotted line):
$I=a \left(V-V_{\text{offset}}\right)$ with $a=1.20$~$\mu$A/V and $V_{\text{offset}}=2.17$~mV.
In addition, the asymptotic fit for large voltages ($a=1.24$~$\mu$A/V and $V_{\text{offset}}=4.57$~mV) is shown (red dashed line).
Note that the experimental result is digitized from reference \cite{Kurdak1998}.}
\label{fig:Kurdak1998_fitting}
\end{figure}

We next compare the asymptotic result with the experimental result \cite{Kurdak1998} measured by Kurdak et al. 
Because the physical parameters are known, this experimental result is suitable for testing the asymptotic equation. 
Figure \ref{fig:Kurdak1998_fitting} shows the experimental (sample A in reference \cite{Kurdak1998}) 
and the asymptotic results for the following experimental conditions from reference \cite{Kurdak1998}: 
$N_{x}=N_{y}=40$ (i.e.,~$\delta=1$), $C_{g}=1.38$~fF, $C=0.25$~fF, $R_{\text{t}}=810$~k$\Omega$, and $T=20$~mK.
The temperature is sufficiently small for neglecting the exponential dependence in eq.~(\ref{IVrelation_asy_2D}),
and we neglect the square of $\varepsilon=0.0181$.
The asymptotic results typically describes the experimental result, as shown in Fig.~\ref{fig:Kurdak1998_fitting}.
The asymptotic fitting parameters lead to $R_{\text{t}} = 789$ k$\Omega$ and $C=0.0128$~fF.
The resistance $R_{\text{t}}$ is in good agreement with the value in reference \cite{Kurdak1998}.
On the other hand, the capacitance $C$ evaluated from the asymptotic equation is smaller than that estimated by Kurdak et al.
Several reasons can be considered.
First, our expression of $V_{\text{offset}}$ neglects higher terms of $\varepsilon$ 
and one might consider the higher-term effects.
Second, the capacitance ratio $\varepsilon$ is sensitive to $V_{\text{offset}}$.
As shown in Fig.~\ref{fig:Kurdak1998_fitting},
the order of $C$ obtained from fitting is different from that estimated by Kurdak et al.,
while the difference of $V_{\text{offset}}$ is 2.4~mV.
In addition, it is difficult to experimentally evaluate the value of $C$ in general.
In fact, Kurdak et al.~stated "our knowledge of $C$ is less precise," 
and they estimated the value of $C$ from the specific capacitance.
Instead, the asymptotic equation may allow us to approximately evaluate the configuration 
of the array and some physical variables ($N_{x}$, $N_{y}$, $R_{\text{t}}$, $C$ and $C_{g}$) 
in experiments from observations of the offset voltage $V_{\text{offset}}$ and the asymptotical slope at large voltages.

% Comparison of asymptotic equation with the experiment result

%%%%%%%
\section{Summary} \label{sec:summary}
We conducted MC simulations to investigate the $I$--$V$ properties of CB arrays. 
To understand the $I$--$V$ property,
our strategy was dividing it into three regions characterized by the path flow of electrons,
and we pay attention to the size (i.e., aspect ratio $\delta$) dependence.
Our main results were
(i)~power-law behavior of the average CB threshold $\bar{V}_{\text{th}}^{(\text{CB})}$,
(ii)~the power-law exponent $\zeta$ in the intermediate voltage region, 
and (iii)~an asymptotic $I$--$V$ curve at large voltages.

We derived an analytical relationship [eq.~(\ref{Vth_width_Nh001})] for the average CB threshold for $N_{x}=1$, 
and we found that the average CB threshold obeys a power-law decay as a function of the aspect ratio $\delta$.
The coefficient $c$ is in agreement with the previous study by MW \cite{Middleton1993}.
In addition, its power-law exponent $\gamma$ is inversely proportional to the longitudinal size $N_{x}$ (i.e.,~$\delta$ at fixed $N_{x}$) [eq.~(\ref{gamma_power_law})].
It is difficult to obtain the analytical form for $N_{x}>1$ 
because trajectories of electrons meander.
Nevertheless, our analytical and simulation results provide a hint for further development of the study.

The size dependences of the exponent $\zeta$ were shown for different array configurations such as SL, TL-l, and TL-z.
The exponent $\zeta$ in arrays of large $\delta$ was considered to be constant so far.
However, we revealed that $\zeta$ logarithmically increases as $\delta$ increases for both the simple and triangular lattices.
Namely, in addition to the array configuration and array dimension, the aspect ratio $\delta$ is a significant variable for discussing the exponent $\zeta$.

We extended the asymptotic equation for 1D arrays without interaction \cite{Bascones2008} 
to 2D arrays with first-order perturbation of the interaction range $\varepsilon$ [eq.~(\ref{IVrelation_asy_2D}) with eq.~(\ref{phi_th_epsilon})]. 
At sufficient large voltages, the equation adequately describes the Ohmic behavior
and the combined tunneling-resistance $R_{c}$ is inversely proportional to $\delta$.
The offset voltage $V_{\text{offset}}$, included in the asymptotic equation, 
differs from the CB threshold $V_{\text{th}}^{(\text{CB})}$.
Instead, the offset voltage can be regarded as the maximum of the $V_{\text{th}}^{(\text{CB})}$ in the limit of $\varepsilon \to 0$ \cite{Narumi_DistVth_2011}.
These asymptotic property well agrees with simulation and experimental results.
Our extended equation allows to estimate physical values and array configuration 
which are experimentally hard to obtain.
The asymptotic equations for other configurations are also expected to show similar results
and to converges to the Ohm's law.
The details will be discussed elsewhere.

%%%%%%%%%%%%%%%%%%%%%%%%%%%%%%%%%%%%%%%%%%%%
\begin{acknowledgment}
We thank Profs.~Takuya Matsumoto, Megumi Akai, and Takuji Ogawa at Osaka University for fruitful discussions. 
This work was partially supported by the MEXT, 
a Grant-in-Aid for Scientific Research on Innovative Areas~"Emergence in Chemistry"~(Grant No.~20111003)
and another Grant-in-Aid for Scientific Research (Grant No.~21340110).
\end{acknowledgment}

%%%%%%%%%%%%%%%%%%%%%%%%%%%%%%%%%%%%%%%%%%%%
\appendix
\section{Total Energy} \label{appendix:total_energy}
The charge of the i-th island is represented by
\begin{equation} Q_{\text{i}}=\sum_{\text{j}}C_{\text{ij}}(\Phi_{\text{i}}-\Phi_{\text{j}})+\sum_{\mu = +, -, g}C_{\text{i},\mu}(\Phi_{\text{i}}-\Phi_{\mu}). \label{charge_relation} \end{equation}
Assuming that the capacitances are nonzero only between neighboring island--island and island--electrode pairs,
eq.~(\ref{charge_relation}) reduces to
\begin{equation} Q_{\text{i}} =\sum_{\text{j}}M_{\text{ij}}\Phi_{\text{j}}-\sum_{\mu = +, -, g}C_{\text{i},\mu}\Phi_{\mu}, \end{equation}
where $M_{\text{ij}}$ denotes the matrix of capacitances defined by eq.~(\ref{def_capacitance_matrix}).
The capacitance $C_{\text{ii}}$ should be zero by definition. 
Equation (\ref{def_capacitance_matrix}) thus indicates that the diagonal elements $M_{\text{ii}}$ are the sum of all capacitances associated with an island, 
and the off-diagonal elements $M_{\text{ij}}$ (i$\ne$j) are the negative of the capacitance between the i-th and the j-th islands. 
The potential $\Phi_{\text{i}}$ is formally solved to obtain
\begin{equation} \Phi_{\text{i}}=\sum_{\text{j}}M^{-1}_{\text{ij}}Q_{\text{j}}+V^{(\text{ext})}_{\text{i}}, \end{equation}
where $V^{(\text{ext})}_{\text{i}}$ denotes the potential corresponding to the electrodes defined by
\begin{equation} V^{(\text{ext})}_{\text{i}}:=\sum_{\text{j}}\sum_{\mu = +, -, g}M^{-1}_{\text{ij}}C_{\text{j},\mu}\Phi_{\mu}. \end{equation}
The total electrostatic energy of the system is equivalent to the sum of the work for storing charge $Q_{\text{i}}$ under potential $\Phi_{\text{i}}$ in each island and the energy of the electrodes, represented as,
\begin{equation} E=\frac{1}{2}\sum_{\text{i, j}}Q_{\text{i}}M^{-1}_{\text{ij}}Q_{\text{j}}+\sum_{\text{i}}Q_{\text{i}}V^{(\text{ext})}_{\text{i}}+\sum_{\mu = +, -, g}Q_{\mu}\Phi_{\mu}, \label{total_energy} \end{equation}
where the last term denotes the energy of the electrodes and $Q_{\mu}$ the charge at the electrodes $\mu \in\{+, -, g\}$. 
Note that the interparticle electrostatic energy must not be double-counted. 

\section{Energy Change} \label{appendix:energy_change}
Let us consider the tunneling of an electron whose charge is $-e$ from the n-th to m-th island. 
The energy change is 
\begin{equation} \Delta E_{\text{n}\to \text{m}}:=\Delta E^{(\text{p})}_{\text{n}\to \text{m}}+\Delta E^{(\text{ext})}_{\text{n}\to \text{m}}, \end{equation}
where $\Delta E^{(\text{p})}_{\text{n}\to \text{m}}$ and $\Delta E^{(\text{ext})}_{\text{n}\to \text{m}}$ denote the energy changes with respect to the first and second terms of eq.~(\ref{total_energy}), respectively. 
The charge changes to $Q^{\prime}_{\text{i}}=Q_{\text{i}}+e\delta_{\text{in}}-e\delta_{\text{im}}$ with the tunneling; therefore,
\begin{eqnarray} & & \Delta E^{(\text{p})}_{\text{n}\to \text{m}}  \cr
& & \cr
& = & \frac{1}{2}\sum_{\text{i, j}}(Q_{\text{i}}+e\delta_{\text{in}}-e\delta_{\text{im}})M^{-1}_{\text{ij}}(Q_{\text{j}}+e\delta_{\text{jn}}-e\delta_{\text{jm}}) \cr
& & -\frac{1}{2}\sum_{\text{i, j}} Q_{\text{i}}M^{-1}_{\text{ij}}Q_{\text{j}} \cr
& = & e\sum_{\text{i}}Q_{\text{i}}\left[M^{-1}_{\text{in}}-M^{-1}_{\text{im}}\right] \cr
& & +\frac{e^{2}}{2}\left[M^{-1}_{\text{nn}}+M^{-1}_{\text{mm}}-2M^{-1}_{\text{nm}}\right] \cr
& = & e\left[V_{\text{n}}-V_{\text{m}}\right]+\frac{e^{2}}{2}\left[M^{-1}_{\text{nn}}+M^{-1}_{\text{mm}}-2M^{-1}_{\text{nm}}\right], \end{eqnarray}
where the trivial relationship $M^{-1}_{\text{ij}}=M^{-1}_{\text{ji}}$ is used and the effective potential is introduced as
\begin{equation} V_{\text{k}}:=\sum_{\text{j}}Q_{\text{j}}M^{-1}_{\text{jk}} = \sum_{\text{j}}M^{-1}_{\text{kj}}Q_{\text{j}}. \end{equation}
Similarly, 
\begin{eqnarray} \Delta E^{(\text{ext})}_{\text{n}\to \text{m}} & = & \sum_{\text{i}}(Q_{\text{i}}+e\delta_{\text{in}}-e\delta_{\text{im}})V_{\text{i}}^{(\text{ext})}-\sum_{\text{i}}Q_{\text{i}}V_{\text{i}}^{(\text{ext})} \cr
& = & e\left[V_{\text{n}}^{(\text{ext})}-V_{\text{m}}^{(\text{ext})}\right]. \end{eqnarray}

Next, let us consider the tunneling from the n-th island to an electrode $\mu$. 
Similar to the above discussion, the energy change is represented as
\begin{equation} \Delta E_{\text{n}\to \mu}:=\Delta E^{(\text{p})}_{\text{n}\to \mu}+\Delta E^{(\text{ext})}_{\text{n}\to \mu}+\Delta E^{(\text{electrode})}_{\text{n}\to \mu}. \end{equation}
Given that the charge change is $Q^{\prime}_{\text{i}}=Q_{\text{i}}+e\delta_{\text{in}}$, the energy change is obtained as
\begin{eqnarray} \Delta E^{(\text{p})}_{\text{n}\to \mu} & = & \frac{1}{2}\sum_{\text{i, j}}(Q_{\text{i}}+e\delta_{\text{in}})M^{-1}_{\text{ij}}(Q_{\text{j}}+e\delta_{\text{jn}}) \cr
& & -\frac{1}{2}\sum_{\text{i, j}}Q_{\text{i}}M^{-1}_{\text{ij}}Q_{\text{j}} \cr
& = &  eV_{\text{n}}+\frac{e^{2}}{2}M^{-1}_{\text{nn}} \end{eqnarray}
and
\begin{equation} \Delta E^{(\text{ext})}_{\text{n}\to \mu} = eV^{(\text{ext})}_{\text{n}}\ ,\ \ \Delta E^{(\text{electrode})}_{\text{n}\to \mu}=-e\Phi_{\mu}. \end{equation}
In addition, the following equations hold:
\begin{eqnarray} \Delta E^{(\text{p})}_{\mu \to \text{m}} & = & -eV_{\text{m}}+\frac{e^{2}}{2}M^{-1}_{\text{mm}}\ , \\
\Delta E^{(\text{ext})}_{\mu \to \text{m}} & = & -eV_{\text{m}}^{(\text{ext})}\ , \\
\Delta E^{(\text{electrode})}_{\mu \to \text{m}} & = & e\Phi_{\mu}. \end{eqnarray}

%%%%%%
\section{Offset Voltage $V_{\text{offset}}$} \label{appendix:phi_th}
The configuration matrix $M^{(\text{1D})}_{\text{ij}}$ for a 1D simple array is represented as
\begin{equation}M^{(\text{1D})}_{\text{ij}} = \left\{ \begin{array}{ll}
(1+2\varepsilon)C_{g} & \ \ \text{i}=\text{j} \\
-\varepsilon C_{g} & \ \ |\text{i}-\text{j}|=1 \\
0& \ \ \text{otherwise} \end{array} \right. \end{equation} 
with arbitrary $\varepsilon = C/C_{g}$.
The inverse elements $M_{\text{i},\text{i}}^{(\text{1D})-1}$ ($\text{i}=1,2,\cdots,N_{x}$) and $M_{\text{j},\text{j}+1}^{(\text{1D})-1}$ ($\text{j}=1,2,\cdots,N_{x}-1$) are derived as
 \begin{eqnarray}M_{\text{i},\text{i}}^{(\text{1D})-1}  & = & \Delta_{\text{i}-1}\Delta_{N_{x}-\text{i}}/\Delta_{N_{x}}, \\
M_{\text{j},\text{j}+1}^{(\text{1D})-1}  & = & \epsilon \Delta_{\text{j}-1} \Delta_{N_{x}-\text{j}-1}/\Delta_{N_{x}},  \end{eqnarray} 
where $\Delta_{n}$ denotes the determinant of the configuration matrix for the 1D simple array that contains $n$ islands and $\Delta_{0}=1$.
The offset voltage reduces to
\begin{equation} V_{\text{offset}}=\frac{e}{\Delta_{N_{x}}}\left(\sum_{\text{i}=1}^{N_{x}}\Delta_{\text{i}-1}\Delta_{N_{x}-\text{i}}-\varepsilon \sum_{\text{i}=1}^{N_{x}-1}\Delta_{\text{i}-1}\Delta_{N_{x}-\text{i}-1}\right). \label{phi_th} \end{equation}
Note that the above representation holds for arbitrary $\varepsilon$.

We can approximately obtain the determinant $\Delta_{n}$ for $\varepsilon\ll 1$ as
\begin{eqnarray} \Delta_{n} & = & \left[ (1+2\varepsilon+\mathcal{O}(\varepsilon^{2}))C_{g}\right]^{n} \cr
& & \cr
& = & (1+2n\varepsilon+ \mathcal{O}(\varepsilon^{2})){C_{g}}^{n}. \label{determinant}  \end{eqnarray}
Substituting eq.~(\ref{determinant}) into eq.~(\ref{phi_th}) leads to eq.~(\ref{phi_th_epsilon}). 

%%%%%%%%%%%%%%%%%%%%%%%%%%%%%%%%%%
%%% REFERENCE %%%%%%%%%%%%%%%%%%%%%%%
%%%%%%%%%%%%%%%%%%%%%%%%%%%%%%%%%%

%
%%%%%%%%%%%%%%%%%%%%%%%%%%%%%%%%%%%

\end{document}